# A Critical Review of the Impact of Candidate Copy Number Variants on Autism Spectrum Disorders

Seyedeh Sedigheh Abedini[1,¶], Shiva Akhavan[2,¶], Julian Heng[3], Roohallah Alizadehsani[4], Iman Dehzangi[5,6], Denis C. Bauer [7,8], Hamid Rokny[*, 1]

[1]The Graduate School of Biomedical Engineering, UNSW Sydney, Sydney, NSW, 2052, AU.

[2] Department of Molecular Biology and Genetics, Yeni Yuzyil University, Istanbul, Turkey

[3] Curtin Health Innovation Research Institute, Curtin University, Bentley 6845, Australia

[4] Institute for Intelligent Systems Research and Innovation (IISRI), Deakin University, Victoria, Australia.

[5] Center for Computational and Integrative Biology, Rutgers University, Camden, NJ 08102, USA

[6] Department of Computer Science, Rutgers University, Camden, NJ 08102, USA

[7]Transformational Bioinformatics, Commonwealth Scientific and Industrial Research Organisation (CSIRO), Sydney, Australia

[8]Applied BioSciences, Faculty of Science and Engineering, Macquarie University, Macquarie Park, Australia

¶ Equally contributed

* Corresponding author: e-mail: h.alinejad@mq.edu.au

## Abstract

Autism spectrum disorder (ASD) is a heterogeneous neurodevelopmental disorder (NDD) that is caused by genetic, epigenetic, and environmental factors. Recent advances in genomic analysis have uncovered numerous candidate genes with common and/or rare mutations that increase susceptibility to ASD. In addition, there is increasing evidence that copy number variations (CNVs), single nucleotide polymorphisms (SNPs), and unusual de novo variants negatively affect neurodevelopment pathways in various ways. The overall rate of copy number variants found in patients with autism is 10%-20%, of which 3%-7% can be detected cytogenetically. Although the role of submicroscopic CNVs in ASD has been studied recently, their association with genomic loci and genes has not been properly studied. In this review, we focus on 47 ASD-associated CNV regions and their related genes. Here, we identify 1,632 protein-coding genes and long non-coding RNAs (lncRNAs) within these regions. Among them, 552 are significantly expressed in the brain. Using a list of ASD-associated genes from SFARI, we detect 17 regions containing at least one known ASD-associated protein-coding genes. Of the remaining 30 regions, we identify 24 regions containing at least one protein-coding genes with brain-enriched expression and nervous system phenotype in mouse mutant and one lncRNAs with both brain-enriched expression and upregulation in iPSC to neuron differentiation. Our analyses highlight the diversity of genetic lesions of CNV regions that contribute to ASD and provide new genetic evidence that lncRNA genes may contribute to etiology of ASD. In addition, the discovered CNVs will be a valuable resource for diagnostic facilities, therapeutic strategies, and research in terms of variation priority.

## Introduction

Autism spectrum disorder (ASD) is defined as a highly heterogeneous and complex condition of Neurodevelopmental Disorder (NDD), which is characterized by varied degree of impaired socialization, speech and language developments, the presence of repetitive patterns of activities, and restricted interests [1]. Additionally, there is often a combination of various psychological manifestation associated with ASD, including schizophrenia, sleep disturbance, epilepsy, attention-deficit/ hyperactivity disorders (ADHD), and anxiety and intellectual disabilities (ID) [2, 3]. The global prevalence of ASD is estimated to be about 1 in 100 children. According to Center for Disease Control (www.cdc.gov/ncbddd/autism/data.html, retrieved on 2 August 2022) he ratio among mans is four to five times more common among males [4-6]. ASD development is influenced by a number of genetic factors and environmental variables. The contribution of genetics to the etiology of autism is undoubtedly significant, and including several factors such as single nucleotide polymorphisms (SNPs), changes within non-coding and regulatory regions such as long non-coding RNAs (lncRNAs), common variants, rare or *de novo* mutations, epigenetics elements, and copy number variants (CNVs) [7-11].

Previous studies have shown the importance of CNVs and SNPs for ASD and other diseases [9, 12-19]. Among ASD patients, even within the same locus region, CNVs display a significant clinical diversity, particularly in phenotypic characteristics, severity and comorbidities [20]. CNVs with frequency rate of 10%-20% are known as a major genetic cause of ASD, but only 3%-7% can be detected by conventional cytogenetic methods [6, 21, 22]. The development of chromosomal microarrays (CMA), genome-wide association studies (GWAS), and next generation sequencing (NGS) techniques have revealed a wide range of CNVs associated with ASD. Currently, at least 90 pathogenic CNVs with causative genes and/or deletion or duplication regions, related to ADS have been identified. Most of these CNVs are inherited in familial pattern. However, they can also occur *de novo* or spontaneously [6, 7] Although it is evidence that CNVs play a major role in ASDs, since most CNVs are incompletely penetrant and variable in expression, it can sometimes be challenging to determine the responsible genes associated with ASD. Therefore, the identification of known and unknown loci and candidate genes for ASD is needed to provide further genotype-phenotype information.

The current study focuses on exploring 47 CNV regions that have been recently identified by our team using novel CNV detection toolsets SNATCNV [16] and PeakCNV [23] on a CNV dataset containing 6,479 controls and 19,663 cases samples with ASD-like phenotype. We used KARAJ [24] to download datasets and the supplementary files. All procedures for detection of ASD CNV regions have been described at the previous publication [16]. In this study, we performed an integrative bioinformatics analysis with the goal of enhancing our understanding of autism's genetic background to provide an overview of both known and novel causative genes within the CNV regions and summarizing existing evidence about the influence of detected variation on brain and nervous system (NS) as a risk factor for ASD.

**Table 1. Identified CNV regions**

| Locus | region ID (hg19) | type | #case | #control | P value | #Brain-enriched / total | | #Mice neuro phenotype |
| --- | --- | --- | --- | --- | --- | --- | --- | --- |
| | | | | | | coding | lncRNA | |
| 1p36.32 | chr1:10001-10270615 | Del | 240 | 41 | 5.10E-06 | 57/155 | 44/151 | 19 |
| 1q21.1 | chr1:146113643-147727385 | Dup | 192 | 12 | 2.60E-17 | 5/10 | 5/15 | 0 |
| 1q21.1 | chr1:146113643-147870095 | Del | 244 | 14 | 6.20E-18 | 5/11 | 5/22 | 0 |
| 2q23.1 | chr2:148669357-148995585 | Del | 50 | 1 | 6.70E-06 | 3/3 | 2/2 | 2 |
| 2q37.3 | chr2:242930600-243007359 | Del | 63 | 1 | 1.70E-07 | 0/0 | 0/3 | 0 |
| 2p16.3 | chr2:51105853-51359022 | Del | 160 | 12 | 2.20E-10 | 1/1 | 1/1 | 1 |
| 3q29 | chr3:195745603-197355603 | Del | 73 | 3 | 5.90E-07 | 12/28 | 9/34 | 2 |
| 4q16.3 | chr4:1705715-2073645 | Del | 75 | 0 | 2.60E-10 | 5/11 | 0/4 | 2 |
| 6p25.3 | chr6:259528-339802 | Del | 91 | 0 | 2.00E-12 | 1/1 | 0/0 | 0 |
| 6p25.3 | chr6:259528-339802 | Dup | 92 | 2 | 2.10E-12 | 1/1 | 0/0 | 0 |
| 7q11.23 | chr7:72742064-74142064 | Del | 146 | 3 | 1.50E-15 | 9/26 | 8/25 | 6 |
| 7q11.23 | chr7:72742064-74142064 | Dup | 108 | 5 | 1.50E-11 | 9/26 | 8/25 | 6 |
| 8p23.1 | chr8:10657597-10695288 | Del | 45 | 0 | 2.70E-06 | 0/3 | 0/0 | 0 |
| 8p23.1 | chr8:7268819-7366446* | Del | 54 | 1 | 2.90E-06 | 0/3 | 0/0 | 0 |
| 8p23.1 | chr8:7721060-7752586 | Dup | 50 | 0 | 4.50E-08 | 0/1 | 0/0 | 0 |
| 8p23.1 | chr8:8650757-11629240 | Dup | 72 | 2 | 4.10E-09 | 9/21 | 18/41 | 0 |
| 9q34.3 | chr9:140441351-140936261 | Del | 58 | 0 | 5.80E-08 | 4/8 | 2/4 | 2 |
| 10q23.1 | chr10:81960020-88800020 | Del | 118 | 15 | 3.90E-05 | 15/26 | 9/32 | 3 |
| 11p14.1 | chr11:29635361-31653568 | Del | 69 | 5 | 4.50E-05 | 7/8 | 7/11 | 1 |
| 13q34 | chr13:111699708-115085141 | Del | 66 | 0 | 5.20E-09 | 10/42 | 10/42 | 1 |
| 15q11.2 | chr15:22798636-23088559 | Del | 245 | 32 | 4.20E-09 | 2/4 | 0/5 | 0 |
| 15q11-13.1 | chr15:23094431-29410239 | Dup | 280 | 23 | 2.80E-21 | 13/17 | 19/47 | 5 |
| 15q11-13.1 | chr15:24674739-28514614 | Del | 147 | 19 | 6.60E-06 | 8/10 | 15/35 | 3 |
| 15q13.3 | chr15:30938215-32510863 | Del | 261 | 24 | 6.30E-14 | 5/11 | 3/13 | 2 |
| 15q13.3 | chr15:32452709-32509897 | Dup | 90 | 7 | 7.90E-08 | 1/1 | 0/0 | 1 |
| 16p13.11 | chr16:15248707-16292499 | Del | 171 | 26 | 1.80E-05 | 5/10 | 2/11 | 0 |
| 16p13.11 | chr16:15502499-16292499 | Dup | 288 | 64 | 8.60E-07 | 5/10 | 1/9 | 0 |
| 16p12.2 | chr16:21942499-22462224 | Del | 111 | 5 | 2.58E-12 | 5/8 | 2/6 | 0 |
| 16p11.2 | chr16:28822499-29052499 | Del | 81 | 2 | 9.40E-09 | 1/9 | 0/8 | 0 |
| 16p11.2 | chr16:29517499-30202499 | Dup | 275 | 28 | 5.80E-18 | 15/32 | 4/20 | 4 |
| 16p11.2 | chr16:29517499-30367499 | Del | 373 | 16 | 8.10E-31 | 16/34 | 4/21 | 6 |
| 17p13.3 | chr17:1038189-1673212 | Dup | 100 | 2 | 4.50E-13 | 6/22 | 1/17 | 4 |
| 17p11.2 | chr17:16446387-18380166 | Dup | 87 | 0 | 8.50E-14 | 17/36 | 10/27 | 5 |
| 17p11.2 | chr17:16789275-18299275 | Del | 61 | 1 | 2.70E-07 | 13/31 | 8/23 | 5 |
| 17p13.3 | chr17:2302359-3502250 | Dup | 85 | 5 | 1.90E-08 | 9/13 | 6/12 | 3 |
| 17p13.3 | chr17:911295-2593250 | Del | 65 | 3 | 5.10E-06 | 14/39 | 8/39 | 8 |
| 17q12 | chr17:34815887-34856055 | Dup | 86 | 6 | 6.20E-08 | 0/3 | 0/0 | 0 |
| 17q12 | chr17:34815887-36225887 | Del | 117 | 5 | 2.40E-10 | 5/17 | 5/12 | 1 |
| 17q21.31 | chr17:43704217-44164182 | Del | 86 | 0 | 8.90E-12 | 3/4 | 4/7 | 3 |
| 22q11.21 | chr22:18545000-21745000 | Del | 456 | 9 | 2.30E-47 | 20/54 | 10/37 | 5 |
| 22q11.21 | chr22:19020000-20308799 | Dup | 193 | 25 | 1.60E-10 | 8/32 | 3/17 | 3 |
| 22q11.22 | chr22:21910000-23650000 | Del | 89 | 9 | 6.30E-05 | 10/64 | 5/22 | 3 |
| 22q11.22 | chr22:23046186-23134119 | Dup | 46 | 0 | 1.20E-07 | 0/7 | 0/5 | 0 |
| 22q13.2-33 | chr22:43000056-51224208 | Del | 259 | 40 | 1.60E-07 | 48/113 | 26/109 | 8 |
| Xq28 | chrX:152561384-154917042 | Dup | 127 | 14 | 2.50E-08 | 17/70 | 2/19 | 6 |
| Xp22.33 | chrX:619146-2700156 | Dup | 146 | 22 | 6.00E-07 | 4/15 | 6/11 | 0 |
| Xp22.31 | chrX:6490000-7885155 | Del | 80 | 7 | 3.80E-05 | 2/4 | 0/4 | 0 |

## Results and Discussions

As a result, we identified 31 known genes in 17 CNV regions and approximately 300 novel candidate genes in the 47 ASD-associated CNV regions (**Table 1** and **Figure 1**) that some of them will be discussed in more details in the following sections.

**1p36.22-p36.33 Deletions**

Our analysis identified a significantly interstitial deleted region at 1p36.22-p36.33 locus (chr1:10001-10270615) that spans 10.2Mb and contains at least 306 genes. A variety of clinical manifestations including congenital abnormalities, developmental delay (DD), ID, muscular hypotonia (MH), and seizures have been observed in patients with different size of 1p36 deletions, both terminal and interstitial deletions in 1p36 [25-27]. Shared clinical manifestations were detected in patients with different sizes of 1p36 deletions, making the correlation between genotypes and phenotypes complicated.

We identified 57 coding genes and 44 lncRNAs within the identified region with significantly enriched expression in the brain. As a result of our study concentrating on the possible roles of genes in the neurodevelopmental process and ASD pathogenesis, 19 coding genes with neural system phenotype in mutant mouse model such as *DNAJC11, DVL, VWA1, GNB1, GABRD, PRKCZ* and *PERE* as well as 16 lncRNAs with dynamic expression during iPSC (induced pluripotent stem cell) to neuron differentiation including *RP1-140A9, TP73-AS1, CATG00000045106, CATG00000004170, CATG00000004816* and *CATG00000004823* has been identified [28-30].

Brain-derived tissues expressed genes are the major proportion of known genetic contributors to autism [16, 31]. Among the brain-enriched genes, *DVL1* (disheveled segment polarity protein 1) is a gene with putative link to ASD and is remarkably eliminated in autistic individuals. Previously, Lijam *et al.* reported impaired social interaction behaviors (a characteristic of autism) in a mouse model lacking *Dvl1* [32]. According to our analysis, using the FANTOMCAT database [28], *DVL1* is highly expressed in the NS of the mouse [29, 30], suggesting it may play a role in neurodevelopment and ASD. *RERE* (arginine-glutamic acid dipeptide repeats) is another brain-enriched gene that is significantly deleted in autistic patients [26]. According to studies on zebrafish and mouse models, mutation and haploinsufficiency of *RERE* leads to NDDs, ID, DD and kidney problems [26, 29, 30, 33-35]. These data showed that *RERE* mutation carriers and autistic individuals had some common phenotypes, indicating this gene is a plausible strong contributor to ASD. We also found a significant deletion of lncRNA *RP11-206L10* on 1p36. This brain-enriched lncRNA was also reported as a gene residing in ID-associated CNVs [36], implying its possible function in neurodevelopmental regulation.

**1q21.1 rearrangements**

In this locus, CNVs have been detected in patients with a broad range of developmental defects and neurological disorders, including DD, ID, schizophrenia, and ASD [37]. Our investigation of 1q21.1 locus reveals the presence of two structural abnormalities: 1.6Mb (chr1:146113643-147727385) and 1.75Mb region (chr1:146113643-147870095) that are significantly duplicated and deleted in autistic peoples, respectively.

Our finding indicates that this locus contains 10 brain enriched genes (5 coding genes and 5 lncRNA genes) that all of them are considerably deleted and/or duplicated in ASD patients.

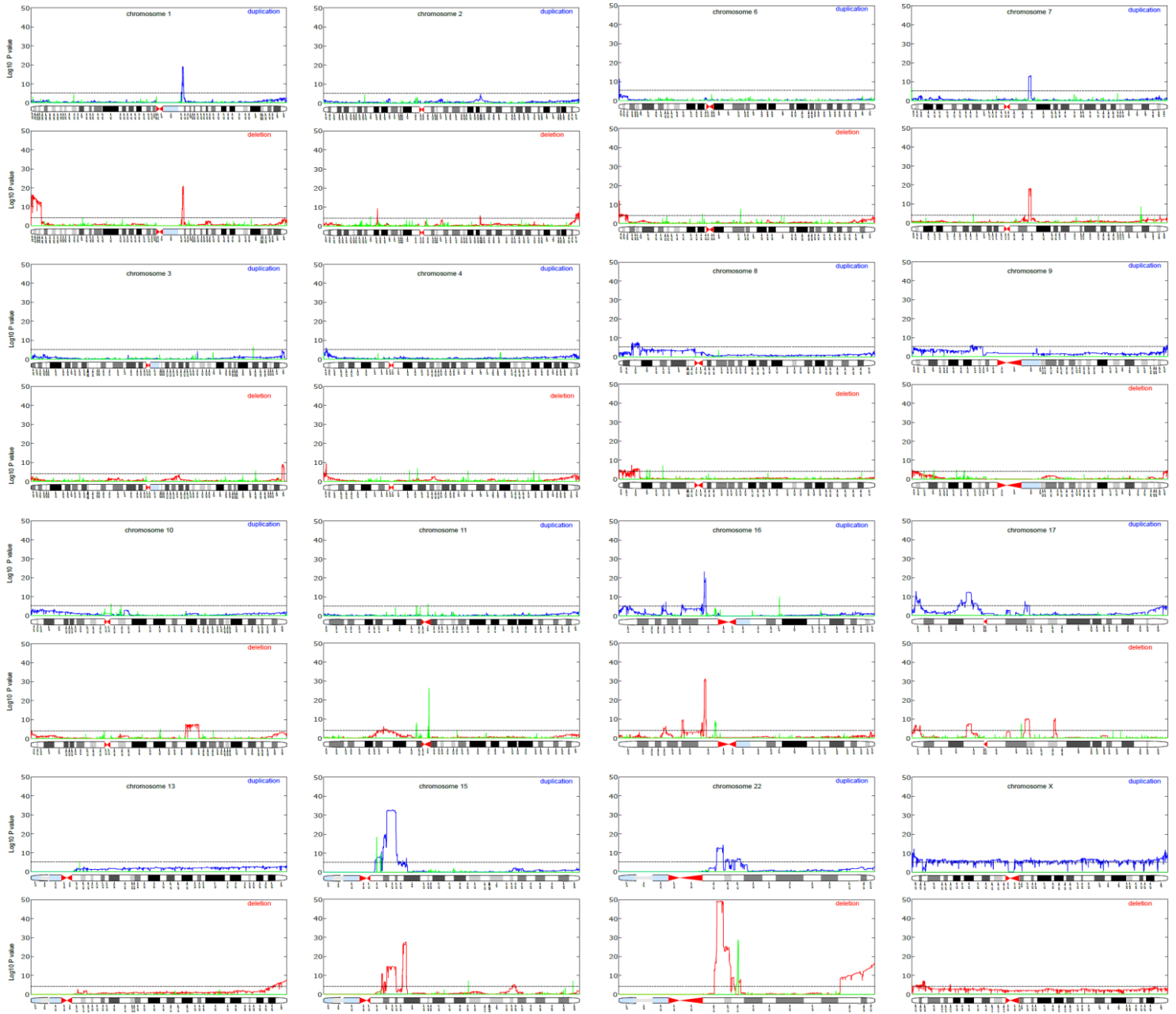

**Figure1. ASD copy number variations on human chromosomes.** Blue: Autistic duplication; Red: Autistic deletions; Green: control copy number variations. Black dashed line indicates our threshold for significant regions (better P-value than 99.9999 percentile of 500,000 permutations). X-axis indicates genomic loci and y-axis indicates logarithm 10 P-value. P-value calculated through one-tailed Fisher exact test.

The 1q21.1 microdeletion have been associated with phenotypes such as Congenital Heart Disease (CHD) [38], schizophrenia [39], ID and microcephaly [27], whereas the 1q21.1 microduplications have been generally associated with ID, macrocephaly, and ASD [27, 37, 40]. The result of our study indicates that brain-enriched *RP11-337C18*, a significantly duplicated/deleted lncRNA with dynamic expression in an iPSC model of neuron differentiation, has potential role in neurodevelopmental process and is likely to be important in ASD. Furthermore, *BCL9* (B-cell CLL/lymphoma 9) as one of brain enriched coding gene within this locus, plays a critical role in signal transduction of canonical Wnt pathway through promoting β catenin's transcriptional activity [41, 42]. The results of association studies have been indicated that Wnt/β catenin signaling dysfunction in neurons may contribute to ASD phenotypes [42, 43]. Also, it has been demonstrated that *BCL9* modulates Wnt signaling through its interaction with *ARX* gene (a causative gene for autism) [44]. This evidence is consistent with our results, suggesting that *BCL9* gene may be involved in ASD.

### 2p16.3 Microdeletion

A 253kb region (chr2:51105853-51359022) on 2p16.3 was significantly deleted in ASD individuals. There are one protein-coding genes *NRXN1* and lncRNA *AC007682.1* in this region, both of which showed brain enrichment, while *NRXN1* had also physiological and morphological effects on mutant mice [29, 30]. *NRXN1* (neurexin 1) a member of the neurexins family of presynaptic cell-adhesion proteins that play a role in synaptic connections, has already been linked to ID and DD. In addition, this gene has been implicated in the pathogenesis of psychiatric disorders such as schizophrenia and autism with incomplete penetrance and variable expressivity of mutations [45-48]. Furthermore, we discover only one NS-enriched lncRNA *AC007682.1* in this region, suggesting a probable role in autism pathogenesis.

### 2q22.3-q23.1 Microdeletions

In autistic individuals, we discover a 326kb region at 2q23.1 with 5 brain-enriched genes (3 coding genes and 2 lncRNA genes) that are significantly deleted. Several phenotypes and disorders are caused by 2q23.1 microdeletions ranging in size from 0.2Mb to 5.5 Mb, such as dysmorphic features, ID, seizures, behavioral problems, and ASD [49-52]. *MBD5* and *ACVR2A* are two of three brain enriched coding genes within the identified region that showed dynamic expression from iPSCs to neuronal differentiations and caused morphological changes in mutant mice [29, 30]. *MBD5* (methyl-CpG binding domain protein 5) belongs to the methyl CpG-binding domain protein family, which encompasses *MECP2,* an already known ASD-associated gene that is mutated in Rett syndrome [51, 53, 54]. It has been proposed that *MBD5* is a dosage-sensitive essential gene in normal development, with autism-related mutations [52, 55-57]

### 2q37.3 Microdeletion

The 2q37 deletion has been associated with brachydactyly mental retardation syndrome (BDMR), a condition characterized by ID, DD, behavioral abnormalities, and ASD [58, 59]. The 2q37.3 microdeletion has also been linked to autism [60] and miscarriage [61]. These findings point to the presence of potentially important regulatory genes in this region, with possible roles in the progression of NDDs. Our analysis identified a 76.76kb as a significantly deleted region at 2q37.3 (chr2:242930600-243007359) in autistic individuals, containing three lncRNA genes (*AC131097.3*; *AC093642.3*; *AC093642.4di*), possibly valuable to be investigated for their involvement in the ASD.

### 3q29 Microdeletion

There are 0.8Mb to 2.1Mb microdeletions in 3q2 region, which are associated with ID, neuropsychiatric and neurocognitive disorders as well as microcephaly, schizophrenia, bipolar disorder, and ASD [62-67]. In the deleted region, both sides are flanked by low copy number repeats (LCRs), suggesting non-allelic homologous recombination (NAHR) at deletion breakpoints [64]. Using SNATCNV, we detected one significantly deleted region (chr3:195745603-197355603) at 3q29 locus which includes 21 brain enriched gene (12 coding and 9 lncRNA genes) out of 64 genes. Among these genes, *FBXO45, PAK2* and *DLG1* were shown to have impact on mutant mouse's NS [29, 30], and *CATG00000067412*, *MFI2-AS1* and *AC128709.2* are three lncRNA genes implicated in iPSC to neuron differentiation with brain enrichment, implying their possible function in neurodevelopmental processes regulation.

*PAK2* (p21 (RAC1) activated kinase 2) is a serine/threonine kinase that regulates actin cytoskeleton dynamic, with potential signaling roles in cortical neurons [68-70]. It is highly expressed in the fetal brain and may contribute to neuronal differentiation [69, 70]. It was also documented that *PAK2* plays a crucial role in the processes of brain development and autism progression [64, 71].

*FBXO45* (F-box protein 45) is an ubiquitin ligase involved in the growth and formation of synapses [72, 73] and that also contributes to the progression of schizophrenia and autism [74]. In addition, studies in mice suggested roles for *Foxo45* in neuronal migration and development [75]. Overall, these findings and genotype-phenotype correlation study for 3q29 deletion indicate that *FBX045* might be a candidate gene involving in the progression of neuropsychiatric disorders [63].

*DLG1* (discs large MAGUK scaffold protein 1) is a paralogue of *DLG3*, a gene engaged in the pathogenesis of ID, when mutated [64]. This gene is required for normal development and is a component of a core transsynaptic complex (Neurexin/Neuroligin/DLG/SAPAP/Shank) whose mutations probably increase autism's susceptibility [74, 76, 77].

### 4q16.3 Microdeletion

The 4q16.3 deletion has been reported in patients with Wolf-Hirschhorn syndrome (WHS) [78]. We identified 15 genes (coding and non-coding) in a significant deleted 367kb region on 4q16.3 (chr4:1705715-2073645) of which five coding genes were expressed highly in the brain. Two of these genes, **FGFR3** and **NAT8L** appear to be potential candidates for neurodevelopmental involvement since both express highly in NS and exhibit phenotypic manifestations in mutant mice [29, 30]. The gene ontology analysis indicates that *FGFR3* is strongly associated with seizures and congenital abnormalities [79-81], which are very common in ASD as well as STAT cascade with roles in neural development [82]. This finding suggests a possible role for *FGFR3* in autism.

### 6p25.3 rearrangements

Our data scanning for finding substantial CNVs related to ASD revealed a novel 80.3Kb region (chr6:259528-339802) on 6p25.3 chromosomal band which is significantly deleted and duplicated. The region contains a single gene **DUSP22** (Dual specificity phosphatase 22), which has previously been revealed to contribute to autism pathogenesis through JNK signaling pathway [83-86]. JNK signaling is believed to have a crucial role in the development of psychiatric diseases such as ID, schizophrenia and ASD [84]. The same study, in parallel to other findings [87], also indicated that *DUSP22* is an imprinted gene expressed maternally in adult prefrontal cortex [86], and further suggested as a possible contributor in the pathogenesis of ASD. In addition, heterozygous deletion of *DUSP22* gene has also been reported in ASD [88].

### 7q11.23 rearrangements

We also identify deletions and duplications of 1.5 to 1.8 Mb at 7q11.23 locus, that results in Williams-Beuren Syndrome (WBS; MIM 194050) and 7q11.23 duplication syndrome (Dup7; MIM 609757), respectively [27, 88-90]. WBS has a 6-10 times higher prevalence in autism compared to the general population [88], indicating that its deletion may play a crucial role in autism pathogenesis. In autistic individuals, we found a significantly deleted/duplicated region (chr7:72742064-74142064), encompassing 51 genes (26 coding and 25 lncRNA genes), spanning 1.4Mb on 7q11.23 chromosomal band. After narrowing the list of significantly deleted/duplicated brain-enriched genes (9 coding and 8 lncRNA genes) located at 7q11.23 locus, we identified *STX1A*, *GTF2I*, *GTF2IRD1*, *CLIP2*, *FZD9*, *LIMK1*, *EIF4H* that are associated with NS phenotype in mutant mouse and *CATG00000096069*, *CATG00000092613*, *CATG00000096089* that display dynamic expression in iPSC model as genes with possible importance in ASD. Furthermore, our analysis of *STX1A* suggests a role for this gene in autism, since it is highly expressed in brain tissues, has significant deletions and duplications in ASD, and shows effects on the NS when it is mutated in mice.

**STX1A** (syntaxin 1A) modulates serotonin transporter activity, which is overexpressed in autism with high functioning and can potentially be involved in the pathogenesis of ASD [91-93]. **GTF2I** (the general transcription factor 2I) and **CLIP2** (CAP-Gly domain containing linker protein 2) are also other NS-enriched genes that may play a role in the etiology and manifestation of autism [94, 95].

### 8p23.1 rearrangements

We also study various deletions and duplications in 8p23.1 as an extremely susceptible locus for genomic rearrangements. Various changes in the range of 3.6 Mb to 6.5 Mb residues were reported by different groups [96-100]. Most interstitial deletions of 8p23.1 are *de novo* [100, 101], whereas terminal deletions are inherited [100]. We narrowed down the regions of CNVs and uncovered four regions between 38kb and 3 Mb that were significantly deleted (chr8:10657597-10695288, chr8:7268819-7366446) or duplicated (chr8:7721060-7752586, chr8:8650757-11629240) in autistic individuals on 8p23.1. Our study revealed that deletions in these regions are associated with an increased risk of NDDs, psychiatric conditions, abnormality of the heart, hyperactivity, and mild to moderate ID [27], whereas duplications in these loci are linked to abnormal facial shape, behavioral/psychiatric abnormality, problems with learning/speaking, malformation of the heart, and ID [27]. In these four regions, we only detect 27 remarkably duplicated brain-enriched genes (9 coding and 18 lncRNA genes). Despite the absence of any reported ASD associated gene in these regions, our study points to a number of significant candidate genes that are either brain-enriched such as *TNKS, PINX1, MSRA, XKR6, FAM167A, NEIL2* or associated with NS phenotypes in mutant mouse like *RP1L1* and *GATA4* [28-30]. In addition, we identified that lncRNA RP11-1E4.1 significantly expressed in brain and dynamically expressed during iPSC to neuron differentiation, suggesting its effect on neurodevelopment and possible contribution to ASD phenotypes.

*MSRA* (methionine- sulfoxide reductase) is located in a remarkably duplicated region on chr8:8650757-11629240. It has a regulatory role in human fetal brain development [102] and protective function against oxidative stress [103], as a causative mechanism for ASD [104]. Furthermore, *MSRA* is associated with bipolar disorder [105] and schizophrenia [106]. All these findings along with our result emphasis the possibility of a strong correlation between *MSRA* and autism.

*PINX1* (PIN2/TRF1-interacting telomerase inhibitor 1) is a regulator of telomere integrity that is required for chromosomal stability [107, 108]. Multiple lines of evidence imply that some genetic variations involved in ASD are linked to telomere shortening [109, 110]. As a result, *PINX1* may be a potential candidate gene for telomere shortening in ASD patients.

*TNKS* (tankyrase 1) is another brain enriched gene in 3Mb duplications region (chr8:8650757-11629240) with suggestive role in ID [28, 100]. This gene regulates sister chromatid separation [111], intestine organoid growth [112], and Wnt/β catenin signaling [113]. Since, both activation and inhibition of canonical Wnt pathway seem to increase autism risk [42], *TNKS* could be considered as a ASD related gene.

**9q34.3 Deletions**

Several types of rearrangement occur at 9q34.3 locus including terminal and interstitial deletions, unbalanced translocations, and complex rearrangements, which are associated with 9q34.3 microdeletion syndrome [114]. These microdeletions with sizes typically less than 3.5-4 Mb and multiple breakpoints at the telomeric regions cause ID, microcephaly, MH, protruding tongue, heart impairment, congenital defects, and developmental impairment in fetuses [27, 114]. Our study introduced a 495 kb region at the 9q34.3 locus (chr9:140441351-140936261) containing 12 genes (8 coding and 4 lncRNA genes), which was significantly deleted in ASD. Among the identified genes, *CATG00000107435* and *CATG00000107433* were two considerably deleted NS-enriched lncRNAs, whereas *CACNA1B* and *EHMT1* were brain enriched coding genes affecting the NS of mutant mice [29, 30].

*EHMT1* (euchromatin histone methyltransferase 1) is reported as a genetic cause of ID syndrome associated with subtelomeric deletion of 9q34.3, due to its haploinsufficiency [115, 116]. This gene has been implicated in the appearance of autistic-like symptoms in knockout mice [117] as well as the development of psychological problems, exclusively mood disorders and ASD in mosaicism status [118].

*CACNA1B* (calcium voltage-gated channel subunit alpha B) belongs to the calcium channel genes family, whose members modulates neural function and possibly contributing to ASD [119, 120]. According to gene ontology analysis, *CACNA1B* causes neurological speech impairment, which is one of the common problems in ASD. Altogether, these findings and our results may indicate a possible involvement of this gene in the progression of ASD.

**10q22-q23 Deletions**

It has been demonstrated that 10q22.3q23.2 deletions are associated with cognitive and behavioral abnormalities as well as autistic features [121]. The 10q23 deletions affect several genes with neurodevelopmental function, including *NRG3*, *GRID1*, *BMPR1A*, *SNCG*, *GLUD1*, and many others [122]. We discovered a significant deleted 6.84 Mb region (chr10:81960020-88800020) in autistic patients using bioinformatics analysis, encompassing 26 coding genes and 32 lncRNA genes, 24 of which (15 coding and 9 lncRNA genes) are brain-enriched. Furthermore, the FANTOMCAT database revealed that *CATG00000000247*, *CATG00000000263*, *RP11-77P6.2*, and *RP11-96C23.5* are highly expressed during iPSC to neuron differentiation [28] as well as *GRID1, OPN4*, and *SNCG* represented neural system phenotypes in mutant mouse [29, 30], suggesting their potential involvement in neural development.

*GRID1* (glutamate receptor ionotropic, delta-1) is suggested to be associated with schizophrenia [123] and contributes to the genetics of autism [124-126]. Additionally, **BMPR1A** (bone morphogenetic protein receptor type-1A) is a significantly deleted gene in patients with ASD, and Juvenile Polyposis Syndrome (JPS) results from loss of function mutations [127]. Microdeletions in *BMPR1A* are associated with severe polyposis and childhood malignancies as well as some autism related phenotypes such as cardiac abnormalities, macrocephaly and DD [128]. Another gene in this region, **NRG3** (neuregulin 3), as a brain enriched gene, is thought to be a schizophrenia susceptibility gene [129], and 10q22-q23 microdeletions including this gene have been linked to NDDs such as cognitive impairment, DD, and autism [129, 130].

## 11p14.1-p13 deletions

There is evidence that the 11p14.1 microdeletion is associated with obesity and autism as well as severe developmental delays, and further clinical features in individuals carrying larger deletions [131]. Using SNATCNV, we found a significantly deleted region spanning 2 Mb (chr11:29635361-31653568) on 11p14.1-p13 loci which contained 19 genes (8 coding and 11 lncRNA genes), where 14 of them were NS-enriched genes (7 coding and 7 lncRNA genes). Among these genes, we identified *KCNA4* (potassium voltage-gated channel subfamily A member 4) as the most significantly deleted gene with an effect on the mouse NS, when mutated [29, 30]. This gene modulates the time interval and shape of the cardiac action potential and heart rate [132] as well as neurotransmitter release [133, 134]. Despite lack of studies that links *KCNA4* and autism, it is known that dysfunction of neurotransmitter system and altered heart rate are linked to autism [135-138], which may be suggestive of a role for *KCNA4* in ASD. *ELP4* (elongator acetyltransferase complex subunit 4) is also another NS-enriched gene, residing in the 11p14.1-p13 CNVs. Deletion of *ELP4* has been demonstrated to be likely principal for the progression of various NDDs, ranging from language impairment to epilepsy and ASD [139].

## 13q34 Deletion

The 13q34 deletion has been associated with high risk for NDDs including epilepsy and mild ID as well as DD [140-142]. Our bioinformatics analysis of CNVs related to ASD, using SNATCNV, revealed a significantly deleted 3.38 Mb region at 13q34 (chr13:111699708-115085141) in autistic patients, encompassing 42 coding genes and 42 lncRNA genes. Among the significantly deleted genes with enrichment in the brain and NS: *CATG00000016319, LINC00403, CATG00000018119, CATG00000018133, CATG00000016366,* and *CATG00000018138* displayed dynamic expression in neural differentiation of iPSCs as well as *ATP11A* that showed to affect the mice's NS, when mutated [29, 30], possibly reflecting their importance in the progression of NDDs.

Deletion of 13q34 harboring *ATP11A* as well as *MCF2L*, *F7*, *F10*, *PROZ*, *PCID2*, *CUL4A*, and *LAMP1* genes within our defined significantly deleted region was detected in an Italian family suffering bleeding diathesis, multiorgan involvement, and language delay, suggesting underlying roles for these genes in ASD-related manifestations such as language delay [143]. In addition, *UPF3A*, as another brain enriched gene that is significantly deleted in ASD, plays a compensatory role in patients with loss of *UPF3B* function [144]. *UPF3B* is a paralog of *UPF3A*, with known implications in ID, childhood onset schizophrenia, ADHD, and autism [145].

## 15q (15q11.2; 15q11.2-q13.1; 15q13.3) rearrangements

The proximal long arm of chromosome 15 contains a cluster of low copy DNA repeats (called duplicons) that facilitates the occurrence of NAHR during meiosis, and five breakpoints known as BP1-BP5 [146, 147]. As a results of these rearrangements, various NDDs including Prader-Willi syndrome (PWS), Angelman syndrome (AS), and ASD has been reported [148-150]. Furthermore, a range of phenotypes and manifestations have been documented in patients with 15q11.2 deletions as well as carriers of 15q11.2-q13.1 rearrangements including electroencephalogram (EEG) abnormality, ID, microcephaly, seizures, schizophrenia, truncal ataxia, hypogonadism, MH, ASD, and truncal obesity [27, 151, 152]. As well as behavioral, cognitive and language impairments and ASD have also been associated with larger deletions at 15q11-q13 [153] and facial abnormalities, ID, and seizures have been linked to 15q13.3 deletion [27]. Further, our bioinformatics analysis detected some relevant 15q CNVs in autistic individuals, including 15q11.2 deletion (chr15:22798636-23088559), 15q11.2-q13.1 duplication (chr15:23094431-29410239), 15q11.2-q13.1 deletion (chr15:24674739-28514614), 15q13.3 duplication (chr15:32452709-32509897) and 15q13.3 deletion (chr15:30938215-32510863).

A number of ASD-related genes have been identified in these regions, such as *UBE3A*, *GABRB3*, *GABRA5*, *GABRG3*, *HERC2*, *APBA2* [154-161] located on 15q11.2-q13.1 together with *ARHGAP11B*, *FAN1*, *TRPM1*, and *CHRNA7* [162-165] mapped on 15q13.3, a region that harbors several genes indicating incomplete penetrance, variable expression, and dosage sensitivity [166]. In addition to the known ASD-related genes, we defined a number of significantly deleted and/or duplicated genes in ASD CNVs with enrichments in the brain and NS, including *NDN*, *UBE3A*, *GABRB3*, *GABRA5*, *APBA2*, *OTUD7A*, and *CHRNA7* coding genes with effects on the NS of mutant mice [29, 30] as well as *CATG00000024385*, *RP11-484P15*,

*CATG00000022353*, *CATG00000022386*, *CATG00000024462*, and *CATG00000022467* lncRNA genes with dynamic expression during iPSCs differentiation as possible candidate genes related to ASD.

One of the genes located on 15q11.2-q13.1, known as **UBE3A** (ubiquitin protein ligase E3A) is an imprinted multifunctional gene that acts as both ubiquitin ligase and transcriptional co-activator with regulatory roles at synapses [167-169]. Copy number changes in this gene leads to NDDs such as 15q11.2-q13.3 duplication syndrome and ASD [170], which is consistent with our findings that this gene may be involved in ASD. Linkage disequilibrium at *UBE3A* was also reported in families with autism [154]. **GABRB3**, **GABRA5,** and **GABRG3** genes encode gamma-aminobutyric acid (GABA) receptor subunits which play crucial roles in neurodevelopment and deficiencies in their function cause autism and autism-related NDDs [171]. Another gene at 15q13.3 locus that codes for a ligand-gated ion channel protein that mediates signal transduction at synapse is **CHRNA7** (cholinergic receptor nicotinic alpha 7 subunit). The dosage-sensitive of this gene seems to be involved in the cognitive and behavioral manifestations [172, 173]. Furthermore, reduced expression of *CHRNA7* has been reported in the frontal cortex of individuals with Rett syndrome and autism[152]. Finally, **NIPA1** (NIPA magnesium transporter), a significantly deleted gene in autistic individuals, which encodes a magnesium transporter protein responsible for mediating $Mg^{2+}$ uptake, is likely to be implicated in some NDDs including ASD [174]. Van der Zwaag et al. suggested *NIPA1* and *CYFIP1* (another gene located at 15q11.2 CNVs) as autism risk genes with function in axonogenesis and synaptogenesis [175]. This evidence together with our results strongly supports the relevance of these genes to ASD.

**16p (16p13.11; 16p12.2; 16p11.2) rearrangements**

16p is an autism-susceptibility locus with several duplications and deletions [176-178]. The 16p11.2 and 16p13.11 rearrangements have been associated with a wide range of neurological disorders such as seizures/epilepsy, ID, and autism [179-183], while abnormality of the face, ID, macrocephaly, and speech articulation abnormalities have been linked to 16p11.2 and 16p12.2 deletions [27, 179]. In addition, several manifestations including speech articulation abnormalities and microcephaly have been detected in 16p11.2 duplication carriers [179]. Along with these findings, our results derived from bioinformatics analysis also revealed six valuable regions in autistic individuals including two deleted regions at 16p11.2 (chr16:28822499-29052499; chr16:29517499-30367499), which were 0.23 Mb and 0.85 Mb in size, respectively, and a 0.68 Mb duplication at 16p11.2 (chr16:29517499-30202499) as well as a 0.52 Mb deletion at 16p12.2 (chr16:21942499-22462224), a 1 Mb deletion at 16p13.11 (chr16:15248707-16292499) and a 0.8 Mb duplication at 16p13.11 (chr16:15502499-16292499).

As a result, we identify 60 genes (47 coding genes and 13 lncRNA genes) with a high-level expression in NS that 35 of them are significantly deleted and 25 are considerably duplicated in ASD patients. In the following, our study shows that *CATG00000028919* and *CATG00000028920*, the only two significantly deleted and brain-enriched lncRNA genes at the 16p12.2, represent dynamic expression during iPSCs to neuron differentiation. It is also noteworthy that **KCTD13, PRRT2, SEZ6L2, DOC2A, TBX6,** and **MAPK3** are NS expressed genes that have effect on the neural system of mutant mice [29, 30], presumably playing roles in NDDs such as autism. In addition, **TAOK2**, another gene at the 16p11.2 CNV region, is implicated in NDDs due to its essential role in dendrite morphogenesis [184, 185].

**KCTD13** (potassium channel tetramerization domain containing 13), a gene found at 16p11.2 CNVs, has crucial role in the regulation of neurodevelopmental processes [186, 187]. A study on zebrafish and mouse models indicates that *KCTD13* is a regulator of head size phenotype, causing microcephaly and macrocephaly in the case of duplication and deletion, respectively [186]. The same study reported dosage changes of *KCTD13* due to deletion and *de novo* structural alteration are consistent related to autism [186, 188], which supports our findings that this gene may contribute to the progression of ASD. **MAPK3** (mitogen-activated protein kinase 3), another brain-enriched gene, is a member of MAP kinase family which regulates brain growth, synaptic target selection and connectivity [189]. The abnormality in the MAP kinase pathway has been linked to different developmental disorders, including autism [190, 191]. Our investigation also revealed that **SH2B1**, a NS-enriched gene with roles in leptin and insulin signaling, glucose homeostasis regulation and body weight, is deleted significantly in autism, suggesting it as a possible candidate genes at 16p11.2 CNVs for obesity as a symptoms of ASD [192-194].

**17p (17p13.3; 17p11.2) rearrangements**

The short arm of chromosome 17 contains a high density of LCRs, making the region susceptible for NAHR occurrence [195]. The 17p13.3 is a genomic instability and recombination hotspot locus, harboring various deleted and duplicated CNVs, associated with different NDDs and development of epileptogenesis [196, 197].

The deletion of gene rich 17p13.3 locus leads to abnormality of the heart, lissencephaly, microcephaly, midface retrusion, and Miller-Dieker syndrome [27, 198]; while its duplication is associated with split-hand/foot malformation, autistic symptoms, postnatal overgrowth, psychomotor and developmental delay, and hypotonia [199, 200].

Furthermore, the deletion of 17p11.2 locus results in Smith-Magenis Syndrome (SMS), hyperactivity, ID, MH, self-mutilation, short stature, sleep disturbance, and stereotypic behavior [27, 201]; whereas Potocki-Lupski Syndrome, (PTLS), autism, hyperactivity, short attention span, and short stature are clinical manifestations of 17p11.2 duplication [27, 202]. Our bioinformatics analysis to identify other possible genes related to autism in ASD CNVs revealed five significantly deleted or duplicated regions including: two duplications at 17p13.3 that are 635 kb (chr17:1038189-1673212) and 1.2 Mb (chr17:2302359-3502250) in size, a 1.7Mb region of deletion at 17p13.3 (chr17:911295-2593250), a 1.93 Mb region of duplication at 17p11.2 (chr17:16446387-18380166) and a 1.5 Mb region of deletion at 17p11.2 (chr17:16789275-18299275). In total, 259 genes were detected, 92 of which were highly expressed in brain (59 coding genes and 33 lncRNA genes) which among them, 49 genes were significantly duplicated, and others were remarkably deleted. To further narrowing the number of possible autism-related genes in the specified regions, we discovered significantly duplicated *CATG00000030828*, *CATG00000032809*, *RP11-74E22.3*, and *CATG00000030472* lncRNA genes as well as significantly deleted *CATG00000032796* and *CATG00000032809* lncRNA at 17p13.3 CNV regions, that are expressed during iPSC to neuron differentiation, possibly indicating their role in neurodevelopmental processes and the development of ASD. Likewise, we identified 16 coding genes such as *ASPA, TRPV1, RTN4RL1 YWHAE, RASD1, RAI1,* and *PAFAH1B1*, mapped to significantly deleted and duplicated regions, affecting the mouse NS when mutated [29, 30], as candidate genes with probable involvement in ASD.

*TRPV1* (transient receptor potential cation channel subfamily V member 1) is a brain enriched gene located at 17p13.3 that is significantly duplicated. It has been demonstrated that this gene is regulated by *SHANK3* [203], one of the best known autism-implicated genes with function in synaptic formation, maturation, and maintenance [204-206], located at 22q13.3 CNVs. In sensory neurons, activation of *TRPV1* have impact on pain perception [207], whereas abnormal pain sensitivity is typically observed in ASD [208]. Therefore, *TRPV1* and its adjacent gene, *TRPV3*, could be possibly implicated in occurrence of this phenotype in ASD.

*RAI1*, another ASD related gene at the 17p11.2 CNV regions, is a dosage sensitive gene tough to be responsible for the most phenotypes relevant to SMS and PTLS [209, 210]. Moreover, *RAI1* has dynamic expression during iPSCs to neuron differentiation, which along with our analysis suggests its possible implication in autism.

Furthermore, *YWHAE* gene (tyrosine 3-monooxygenase/tryptophan 5-monooxygenase activation protein epsilon) encodes a protein that interacts with key signaling molecules including calmodulin and MAP3K. Duplication of *YWHAE* has been postulated to promote the activity of ERK-pathway, which is inappropriately activated in ASD patients [211-214].

Finally, *PAFAH1B1* (platelet activating factor acetyl hydrolase 1b regulatory subunit 1; also known as lissencephaly gene 1 (*LIS-1*)), located in the 17p13.3 CNV regions, is another candidate ASD related gene due to its dynamic expression during iPSC to neuron differentiation and impact on the neural system of mutant mouse [29, 30]. Additionally, the presentation of lissencephaly in Miller-Dieker syndrome is caused by mutation or deletion in *PAFAH1B1* gene [215], while hypotonia and mild developmental delay have been linked to its duplications [214].

**17q (17q12; 17q21.31) rearrangements**

Speech and motor delay, seizures, vision problems, and behavioral abnormalities are some of disease related to 17q12 recurrent duplication, which usually indicate reduced penetrance and variable expression [216]. The deletion of 17q12 confers high risk for schizophrenia and ASD [217] and are also associated with different diseases such as liver cancer, diabetes mellitus, and multiple renal cysts [27, 218]. In addition, 17q21.31 microdeletion and duplication has been reported to cause Koolen-de Vries syndrome (KdVS), a disorder with various symptoms such as DD/ID, neonatal/childhood hypotonia, dysmorphisms, congenital malformations, speech and language delay [27, 219], as well as frontotemporal lobar degeneration (FTLD) and schizophrenia in a few patients, respectively [220, 221].

Our bioinformatics analysis displays three regions at 17q harboring CNVs that are significantly rearranged in ASD, including a 40 kb region of duplication at 17q12 (chr17:34815887-34856055), a 1.4 Mb region of deletion at 17q12 (chr17:34815887-36225887), and a deleted region of 0.46 Mb in size at 17q21.23

(chr17:43704217-44164182). Eight out of 24 coding genes and nine out of 19 lncRNA genes detected in the significantly duplicated/deleted ASD CNV regions are highly expressed in the brain. A number of these gene may have an impact on neurodevelopment and autism, including *RP11-445F12*, *RP11-697E22,* and *MAPT-AS1* lncRNAs with dynamic expression during iPSC to neuron differentiation as genes as well as *LHX1*, *CRHR1*, *MAPT,* and *KANSL1* with effects on the NS of mutant mouse and significant deletion in ASD [29, 30].

*LHX1* (LIM homeobox 1), mapped to 17q12 locus, is a brain-enriched gene rearranged in various diseases including autism [222]. In addition, [223] suggested the likelihood of involvement of *LHX1* in the development of brain. ***MAPT*** (microtubule associated protein tau) and ***KANSL1*** (KAT8 regulatory NSL complex subunit 1) are two genes located at CNVs in the 17q21.31 region. The *MAPT* gene encodes tau proteins that are involved in Alzheimer's disease (AD) and Parkinson disease [224]. Furthermore, it has been shown that mutations and duplications of *MAPT* are associated with frontotemporal lobar degeneration (FTLD), whilst its deletions is linked to several manifestations like facial dysmorphism, hypotonia, and ID [225]. [226] discovered that the reduction of tau protein prevents autism-like behaviors in mouse models. In contrast, haploinsufficiency of *KANSL,* the chromatin modifier gene, caused 17q21.31 microdeletion syndrome, that is characterized by ID, distinctive facial features, and hypotonia [227].

**22q11.2 (22q11.21 and 22q11.22) rearrangements**

A cluster of eight LCRs (*LCR22A-H*) in the proximal long arm of chromosome 22 facilitates NAHR and results in various rearrangements and expansion of different combinations of recurrent CNVs in the region [228]. The 22q11.21-q11.23 deletion leads to 22q11.2 deletion syndrome (DiGeorge syndrome:DS), that is the most common microdeletion syndrome in humans, occurring in approximately 1/4000 live births and 1/1000 fetus [229]. Individuals with DS showed heterogeneous clinical presentations, psychiatric and learning problems as well as ASD [230-233]. In order to find significant chromosomal regions and ASD-related genes within CNVs at 22q11.21-q11.22 loci in autistic individuals, we determined four important regions in our study.

The identified regions include a significant deletion and duplication with length of 3.2 Mb (chr22:18545000-21745000, containing 91 genes) and 1.29 Mb (chr22:19020000-20308799, consist of 49 genes) at 22q11.21, respectively, as well as a significant deletion and duplication spanning 1.74Mb (chr22:21910000-23650000, including 86 genes) and 87.9 Kb (chr22:23046186-23134119; comprised 12 genes) at 22q11.22. The 22q11.21 deletion with 30 brain-enriched genes is the most remarkably deleted region among our 47 CNV regions. There have been reports that this deletion is associated with cardiac abnormality, delayed speech and language development, hypocalcemia, nasal speech, and T lymphocytopenia [27]. In the significantly duplicated region at 22q11.21, 11 genes with high level expression in NS (8 coding genes and 3 lncRNA genes) have been identified, including genes related to with ID, nasal speech, and telecanthus [27]. The 22q11.22 deletion with 15 NS-enriched gene is associated with face abnormalities, ID, short stature [27].

We further discovered that *CRKL*, *RTN4R*, *ZDHHC8*, *SEPT5*, *MAPK1*, *GNAZ,* and *BCR* with considerable deletion and significantly duplicated *RTN4R*, *ZDHHC8*, and *SEPT5* genes have effects on the NS of mutant mice [29, 30] as well as *XXbac-B444P24.14*, *DGCR5*, *AC000068.5*, *CATG00000058233*, *CATG00000058251,* and *LL22NC03-86G7* with remarkable deletion and significantly duplicated *XXbac-B444P24.14* and *AC000068.5* lncRNA genes dynamically expressed in iPSC model.

*SEPT5* (SEPTIN 5), a brain enriched gene located within ASD CNV region, is a GTPase with cytokinesis activities and regulates neurotransmitter release at synapses. It has been reported that there is a genetic background-dependent relationship between *Sept5* deficiency and ASD-related phenotypes using mouse model [234, 235]. Another gene, *MAPK1* as a member of MAPK family, plays role in proliferation, differentiation, apoptosis, development, and synaptic plasticity [236-238] and its deficiency results in brain impairment and autism-related traits [239-241]. Furthermore, we identified that *CLTCL1* (clathrin heavy chain-like 1) and *GNB1L* (G protein subunit beta 1 like) significantly deleted or duplicated in autistic individuals are potential genes involved in ADS [242, 243]. Along with coding genes linked to ASD, *DGCR5*, a lncRNA gene with putative involvement in schizophrenia [244], has been proposed as a possible candidate autism-related gene.

**22q13.2-q13.33 Deletion**

The 22q13 deletion syndrome also known as Phelan-McDermid Syndrome (PMS) is caused by deletion ranging in size from 100 kb to over 9 Mb at 22q13.2-q13.33 loci. While larger deletions have been associated with various manifestations and neurological symptoms including delayed speech and language development,

DD, macrocephaly, hyperactivity, and ID; smaller deletions (median size of 3.39Mb) rarely cause these phenotype and have been found in individuals with ASD [27, 204]. In this study, we identified a 8.2 Mb region (chr22:43000056-51224208), that is significantly deleted in ASD and encompasses 113 coding genes and 109 lncRNA genes. Among the 48 brain-specific coding genes, *SHANK3, MAPK8IP2, TTLL1, MLC1, BRD1, PANX2, MAPK11,* and *PLXNB2* were identified to impact on the NS of mutant mouse NS [29, 30], that may suggest possible connections between the function of these genes and neurodevelopmental processes. In addition, *CATG00000057830, RP3-388M5, CTA-217C2, CATG00000057926, RP3-402G11.26, CHKB-AS1, CATG00000058090,* and *CATG00000058095* from *CTA-217C2*the 26 brain-enriched lncRNA genes showed dynamic expression during iPSC to neuron differentiation (using FANTOMCAT database [28]), which may indicate their importance in the development of neurological conditions, despite the fact no study has yet demonstrated such correlation.

*SHANK3*, as a deleted gene in PMS, is the most well-known candidate gene for neurological manifestations in 22q13 deletion syndrome and ASD. It plays role in the formation, maturation, and maintenance of synapse [204, 205] and also associated with ASD phenotype [245]. An additional candidate gene for Chr22qter-associated disorders that is almost deleted in PMS and ASD cases [246] is *MAPK8IP2* (also known as *IB2*). It encodes a scaffold protein involved in JNK signaling [247] with high enrichment in postsynaptic densities. Both *MAPK8IP2* and *SHANK3* are brain specific genes that highly expressed during iPSC to neuron differentiation with effects on the NS of mutant mice [29, 30], implying roles for these genes in the development of NDDs including autism. Moreover, *PANX2* is another significantly deleted NS-enriched gene at 22q13.33 locus that has been shown to be differentially expressed gene in autism [248] and has effects on the mouse brain [29, 30].

**Xq28 Duplication**

The Xq28 sub-chromosomal band is a gene rich region (harbors about 13% of the total X-linked genes) which is associated with more than 40 of roughly 300 X-linked diseases [249]. Rett syndrome (a disorder representing autistic phenotypes), Xq28 duplication syndrome, and *MECP2* duplication syndrome have all been reported to be linked to CNVs at this locus [250, 251]. A study using chromosome microarray analysis revealed Xq28 to be the most common subtelomeric locus with copy number gains among patients with DD, ID, dysmorphic features, multiple congenital anomalies, and seizure [252]. Further, it has been documented that delayed speech and language development, severe ID, MH, recurrent infections, seizures, and spasticity are some manifestations associated with Xq28 duplication syndrome [27, 251].

In this study using bioinformatics pipeline, we identified a significantly duplicated 2.35 Mb region (chrX:152561384-154917042) in ASD contains 19 brain-enriched (17 coding genes and 2 lncRNA genes) out of 89 genes. Among the NS specific genes, *MECP2* (already known to be ASD-associated), *GDI1, SLC6A8, L1CAM*, and *PLXNA3* genes with NS phenotypes in mutant mice [29, 30] as well as RP11-66N11.8 lncRNA with dynamic expression in iPSCs model are suggested to be potentially promising candidate as NDD-related genes. Likewise *MECP2* [253-255] and *SLC6A8* [256, 257], and some other genes in the region such as *RAB39B, RPL10, TMLHE,* and *SPRY3* are already known to be associated with ASD [258-260].

*L1CAM* (L1 cell adhesion molecule) and *TMLHE* (trimethyllysine hydroxylase, epsilon) are two candidate risk genes for ASD with role in the development of the NS [261] and transmission of fatty acid through the mitochondrial membrane, respectively [262]. There is evidence that some autistic patients carry rare deletion mutations in *TMLHE* linked to autism [262, 263]. Furthermore, *SPRY3* (sprout RTK signaling antagonist 3), another autism susceptibility gene adjacent to *TMLHE*, showed high expression levels in central and peripheral NS ganglion cells in both mouse and human [264], which is consistent with our findings that suggests *SPRY3* as a highly significant duplicated NS specific gene at 8q28. Another gene that impacts on the cognition function is *GDI1* (GDP dissociation inhibitor 1) [265], which is the most significantly duplicated gene in the ASD region. Considering the link between ID and *GDI1* duplication [251] as well as its effect on the NS of mutant mouse, this this gene may be a strong candidate to contribute to ASD and possibly other NDDs. Finally, *ATP2B3* (ATPase plasma membrane $Ca^{2+}$ transporting 3) with function in $Ca^{2+}$ extrusion, global and local $Ca^{2+}$ homeostasis, and intracellular $Ca^{2+}$ signaling [266], is another brain-enriched gene that significantly duplicated in autistic individuals. It is well known that $Ca^{2+}$ signaling pathways have vital role in the regulation of neuronal development and may be involved in the pathogenesis of ASD [267]. These findings suggest that *ATP2B3* might be a strong candidate gene for ASD.

**Xp22.33 Duplication**

The Xp22.33 spans 4.4 Mb on the short arm of chromosome X and its duplication (Xp22.33p22.32) has been reported with different clinical diseases including ID, DD, delayed speech and language development, short stature, and autism [27, 268]. At this research, we discovered a 2.08 Mb region on Xp22.33 (chrX:619146-2700156) that contains 11 brain enriched genes out of 26 (15 coding genes and 11 lncRNA genes) and is remarkably duplicated in autistic individuals. One of these genes, *ASMT* (acetylserotonin O-methyltransferase) with enzyme activity in melatonin synthesis is located at the pseudoautosomal region (PAR) of the X chromosome. This gene has regulatory effects on sleep, circadian rhythms, and memory with possible influence on cognitive function [269-271]. Moreover, *ASMT* has been identified as a possible autism susceptibility gene due to its rare deleterious mutations and SNP association among ASD patients [272, 273]. In addition, *ASMTL* (acetylserotonin O-methyltransferase-like gene), which is an ASMT-like gene together and ASMTL-AS1 (ASMTL-AS1 antisense RNA 1) lncRNA divergent, with significant duplication in ASD are both located at Xp22.33 locus, implying a possible role for these genes in the progression of neurological disorders and autism. Furthermore, *CATG00000112948* lncRNA is one of the brain-enriched genes in this region that shows dynamic expression during iPSC to neuron differentiation, perhaps highlighting its relevance to NDDs and ASD.

**Xp22.31 Deletion**

The Xp22.31 deletion has been associated with X-linked ichthyosis (XLI), a condition that has been characterized by a variety of symptoms including ocular changes, cryptorchidism, ID, epilepsy, hyperactivity, and autism [274, 275]. We detect a 1.4 Mb region (chrX:6490000-7885155) with significant deletion in ASD, containing four coding genes and four lncRNA genes. Besides *STS*, as causative gene for XLI and various NDDs, *PNPLA4* were expressed highly in the brain. *PNPLA4* (Patatin Like Phospholipase Domain Containing) encodes an enzyme that belongs to the phospholipases of the patatin-like family. In addition to having triglyceride lipase and transacylase activities, the encoded enzyme may play a role in adipocyte triglyceride homeostasis and is associated with human obesity. Recently, it has been suggested that this gene could be implicated in X-linked ID due to its high expression level in the human brain [276]. It is important to note that this finding is in support of our results that *PNPLA4* might play a role in NDDs and ADS.

**Conclusion**

Autism is a neurological disorder whose causes are both genetic and environmental. ASD risk factors continue to be explored in research, and our findings provide a guide for further etiological investigation especially genetics background. Genetic variations have been demonstrated to account for most of the heritability of ASD, among which CNVs are responsible for a significant proportion of ASD cases.

In this study, using SNATCNV tool along with the obtained evidence from bioinformatics analysis and literature search emphasized a strong CNV contribution to ADS. Furthermore, it has been identified that gene disruption can be caused by both deletion and duplication of CNVs. Autism-associated CNVs ultimately affect the development of neurons by impairing gene expression in the brain and have phenotypic impact on NS of mice mutant. As a result of prior studies, approximately 300 candidate genes (coding and non-coding genes) have now been identified to be linked to ASD that most of them were brain-enriched lcnRNA genes. Using iPSCs model of neuron differentiation has showed that approximately 100 lcnRNA were dynamically expressed, indicating that they may play a critical role in neurodevelopment process. In comparison with other studies that have been conducted previously, our findings in this matter are in line with the prior studies [277-279]. As gene expression can be affected by lncRNAs activity in both trans and cis situations [280, 281], there is a possibility that lncRNAs may cause ASD by regulating the expression of protein coding genes involved in this disorder. This study also suggests a list of lncRNAs associated with ASD that can be explored further to determine their potential function. These non-coding genes can be further investigated through chromatin interactions data analyses through publicly available tools such as MaxHiC [282] and MHiC [283].

In addition, this research adds a valuable data to the growing list of CNVs associated with autism by increasing the number of ASD risk gene (coding and non-coding genes), as well as accelerating future genetic studies on ASD including therapeutic pathways and genetics diagnostic tests, based on the system-wide analysis of the putative role of candidate genes (coding and lcnRNA genes) in ASD. Each of the identified CNV and ASD risk genes can be critical for the accurate explanation and personalized intervention of the causative factors for autistic patients due the complicated role of CNVs in ADS etiology.

Furthermore, it has been shown that the SNATCNV is a suitable and time-efficient tools for detection of ASD CNVs, because of its power to reliably discover CNV regions and decrease the genomic space for causative CNV in ASD in a short period of time.

## Materials and Methods

### Samples collection

The SFARI CNV dataset was downloaded through the public SFARI data portal (https://gene.sfari.org/autdb/CNVHome.do). There are 19,663 autistic samples with 47,189 CNVs and 6,479 healthy samples with 24,888 CNVs in this dataset. As CNVs have been reported in different genome build versions such as hg17, hg18, and hg19, we first converted all CNV coordinates to UCSC hg19 using UCSC Lift Genome Annotations tools [284] and confirmed the locations with NCBI remap tools (www.ncbi.nlm.nih.gov/genome/tools/remap). We excluded 4,500 CNVs for which there was no information provided about their genomic coordinates. Further information is provided in the Supplementary table S1 for more details.

### Literature search strategy

Our literature searches were focused on human and mice English language papers available in the PubMed, Scopus, and Web of Science. We also used data and text mining techniques to extract additional related genes [285, 286]. Knowledge-based filtering system techniques have been also used to categorize the texts from the literatures search. The search terms included "Autism", "ASD", "noncoding RNA", "CNV", "copy number variations".

### Identification of significant regions

To identify regions that are recurrently duplicated or deleted in ASD, SNATCNV [16] and PeakCNV [23] mined previously published ASD and control copy number variation databases that are collected and combined by the Simons Foundation Autism Research Initiative (https://gene.sfari.org/database/cnv). For every position in the genome, SNATCNV counted the number of deletion and duplication CNVs from 19,663 ASD patients and 6,479 controls that overlapped and then calculated a *P-value* for each position using Fisher's exact test. To identify significant regions, SNATCNV calculated *P-values* for 500,000 random permutations of case/control labels to estimate the probability that an association emerges by chance. As a result, we identified 47 CNV regions that are significantly deleted or duplicated for ASD case samples (Supplementary table S1).

The 47 CNV regions contain 856 protein-coding genes and 776 non-coding RNAs. We used normalized tag counts from the FANTOM5 expression atlas FANTOM5 [287] to identify nervous system enriched protein coding genes and non-coding RNAs within the 47 CNV regions. To determine whether a gene had enriched expression in the 101 nervous system samples profiled by the FANTOM5 consortium (compared to the total set of 1,829 samples profiled by FANTOM), we first ranked samples by expression and then selected the top 101 samples. We then used Fisher's exact test to determine a *P*-value indicating whether the top 101 samples were more likely to be nervous system samples or not. To determine a threshold on the *P* value we carried out 50,000 permutations randomizing the sample labels and determined the thresholds at the 99% confidence interval. Then, identified any gene with a *P*-value better than this permutation-specific threshold as nervous system-enriched gene [7].

We also annotated protein-coding genes with the information from the Mouse Genome Informatics (MGI) resource (http://www.informatics.jax.org; June 12[th] 2018) to identify those genes that showed a nervous system phenotypes in Mouse mutant (Supplementary table S1). FANTOM5 dataset was also used to identify non-coding RNAs with upregulation in iPSC to neuron differentiation. We then compared all genes in the 47 CNV regions to the 96 known causal genes from MSSNG [288] and SFARI [289]. If a region contained one or more of these genes, we considered it as a region with known ASD genes. We then performed a comprehensive literature search analysis for the remaining CNV regions to check if there are any previously ASD-associated genes in the regions. Our extended literature search was specifically focused on "gene name + autism" or "gene name + ASD". DECIPHER [27], SFARI [289], MSSNG [288] and two largest CNV studies of Global developmental delay to date by Coe *et al.*[290] and Cooper *et al.* [291] were used to annotate the CNV regions.

### Funding

This work was supported by the UNSW Scientia Program Fellowship and the Australian Research Council Discovery Early Career Researcher Award (DECRA) under grant DE220101210 to HAR. The work was also supported by a grant to JI-TH from the Telethon-Perth Children's Hospital Research Fund.

**Acknowledgements**

Analysis was made possible with computational resources provided by the UNSW BioMedical Machine Learning Laboratory (BML) Servers with funding from the UNSW Scientia Program Fellowship. We kindly acknowledge Government of Western Australia, Department of Health, Clinical Excellence, for their support of this project through MERIT award to HAR.

**Author contributions**

HAR designed the study. HAR, SSA, SA, IR, ID, DB wrote and edited the manuscript with help from JH. HAR carried out all the analyses including the statistical analyses, region identification, text mining, gene prioritization, and gene ontology. SSA, SA curated the genes and regions. HAR generated all figures and tables. SA, SSA performed all the literature analysis. All authors have read and approved of the final version of the paper.

**Conflicts of Interest**

The authors declare no competing financial/non-financial interests.